\documentclass[letterpaper, 10 pt, conference]{ieeeconf}  

\IEEEoverridecommandlockouts                              

\usepackage{amsmath}
\usepackage{amssymb}
\usepackage{amsthm}
\usepackage{booktabs}
\usepackage{bm}
\usepackage{etoolbox}
\usepackage{graphicx}
\usepackage{tikz}
\usetikzlibrary{patterns}
\usepackage{pgfplots}
\pgfplotsset{compat=1.18}
\usepackage{xcolor}        
\usepackage[dvips]{epsfig}
\DeclareGraphicsExtensions{.eps,.pdf}
\usepackage[bookmarks=true]{hyperref}
\usepackage[utf8]{inputenc}
\usepackage{siunitx}
\usepackage{tikz}
\usepackage{xfrac}
\usepackage{tabularx}
\usepackage{multirow}
\usepackage{enumerate}
\let\labelindent\relax
\usepackage[shortlabels]{enumitem}
\usepackage{import}
\usepackage{comment}
\usepackage[caption=false,font=normalsize,labelfont=sf,textfont=sf]{subfig}

\newtheorem{theorem}{Theorem}
\newtheorem{remark}{Remark}

\newcommand{\qmain}{\bm{q}^\textup{main}(k)}
\newcommand{\pmain}{\bm{p}^\textup{main}(k)}

\newcommand{\naux}{n^\textrm{aux}}
\newcommand{\qaux}{\bm{q}^\textup{aux}(k)}
\newcommand{\paux}{\bm{p}^\textup{aux}(k)}

\newcommand{\Vfov}{\mathcal{V}^\textup{FoV}(k)}
\newcommand{\obs}{\mathcal{O}}
\newcommand{\nobs}{n^\textrm{obs}}
\newcommand{\qobs}{\bm{q}^\textup{obs}}
\newcommand{\pobs}{\bm{p}^\textup{obs}}

\newcommand{\Vobs}{\mathcal{V}^\textup{obs}}
\newcommand{\nw}{n^\textrm{wp}}
\newcommand{\Ts}{T^\textrm{s}}
\newcommand{\qone}{\bm{q}_1(k)}
\newcommand{\qtwo}{\bm{q}_2(k)}
\newcommand{\qthree}{\bm{q}_3(k)}
\newcommand{\pone}{\bm{p}_1(k)}
\newcommand{\ptwo}{\bm{p}_2(k)}
\newcommand{\pthree}{\bm{p}_3(k)}

\newcommand{\pauxpmain}{\overline{\paux \pmain}}
\newcommand{\ponepobs}{\overline{\pone \pobs}}

\title{\Large \bf
Achieving multi-UAV best viewpoint coordination in obstructed environments}

\author{Mirko Baglioni*$^{1}$, Apurva Patil$^{2}$, Luis Sentis$^{3}$ and Anahita Jamshidnejad$^{1}$ 
\thanks{$^{1}$M.\ Baglioni and A.\ Jamshidnejad are with the Department of Control and Operations, Delft University of Technology, 2629 HS Delft, The Netherlands (*Corresponding author: {\tt\small M.Baglioni@tudelft.nl}).}
\thanks{$^{2}$A.\ Patil is with the Department of Mechanical Engineering, University of Texas at Austin, TX 78712 Austin, United States.}
\thanks{$^{3}$L.\ Sentis is with the Department of Aerospace Engineering, University of Texas at Austin, TX 78712 Austin, United States.}
}

\begin{document}

\maketitle
\thispagestyle{empty}
\pagestyle{empty}

\begin{abstract}
Wildfire suppression is a complex task that poses high risks to humans. 
Using robotic teams for wildfire suppression enhances the safety and efficiency of detecting, monitoring, and extinguishing fires. 
We propose a control architecture based on task hierarchical control for the autonomous steering of a system of flying robots in wildfire suppression. 
We incorporate a novel line-of-sight obstacle avoidance method that calculates the best viewpoints 
and ensures an occlusion-free view for the suppression robot during the mission.
Path integral control generates optimal trajectories towards the goals. 
We conduct an ablation study to assess the effectiveness of our approach by comparing it to scenarios where these key components are excluded, in order to validate the approach in simulations using Matlab and Unity. 
The results demonstrate significant performance improvements,
with $\boldsymbol{44.0 \%}$ increase in effectiveness with the new line-of-sight obstacle avoidance task and up to $\boldsymbol{39.6 \%}$ improvement when using path integral control.
\end{abstract}

\section{Introduction} \label{sec:introduction}
Using robots in severe wildfire is expected to reduce the risks to firefighters' lives and to accelerate the firefighting process. 
Flying robots, or \textit{unmanned aerial vehicles} (UAVs), are particularly promising for wildfire suppression due to their high maneuverability, ability to provide aerial perspectives, and capacity to cover large areas quickly \cite{grogan_pellerin_gamache_2018, 
esteves_et_al_2024}. 
In this paper, we introduce a control architecture for multi-UAV systems to autonomously monitor and extinguish wildfires.
We focus on two categories of UAVs: 
Fire extinguishing UAVs,
which can suppress fire using extinguishing agents and are often teleoperated;
Auxiliary UAVs, that navigate autonomously, supporting the fire extinguishing UAV by 
continuously providing obstacle-free views for human operators. 
Our goal is to fully automate the tasks of the auxiliary UAVs by integrating optimization and reactive control approaches \cite{deKoning_jamshidnejad_2022}.
Auxiliary UAVs might be required to perform multiple tasks, including navigating, providing best views of fire-suppressing UAVs, 
and avoiding collisions. 
These behaviors can be aided using \textit{task hierarchical control} (THC) \cite{slotine_siciliano_1991}, a control approach that enables the parallel execution of multiple tasks with different tracking cost priorities.
In addition, tasks that demand optimality (e.g., closely following a given trajectory to reach a target) can be performed via \textit{path integral control} (PIC) \cite{kappen_2005}, a stochastic optimal control approach which can efficiently be used in real-time. 
In addition, to determine the optimal viewpoints of fire extinguishing UAVs using auxiliary UAVs, we adopt the state-of-the-art best viewpoints method \cite{dufek_et_al_2021}. 
Our main contributions are:
\begin{itemize}
    \item 
    We achieve an occlusion-free field of view (FoV) of auxiliary UAVs for extended periods, with the FoV centered on the fire extinguishing UAV, by introducing a novel line-of-sight (LoS) obstacle avoidance method. This method differs from existing approaches in how it uses the LoS to track targets.
    \item 
    We determine the best viewpoints for auxiliary UAVs by leveraging the state-of-the-art best viewpoints method for cases where obstacles are present, and integrating it with our approach, which combines THC with PIC.
    \item 
    We apply the resulting THC-PIC-Best Viewpoint architecture in a case study focused on fire monitoring and suppression using UAVs.
\end{itemize}

The rest of the paper has the following structure: 
Section~\ref{sec:background} gives the related background; 
Section~\ref{sec:methodologies} describes our proposed methods; 
Section~\ref{sec:simulations} gives the results of a case study that will be discussed in Section~\ref{sec:discussion}; 
Section~\ref{sec:conclusion} concludes the paper and gives topics for future work.
Moreover, Table~\ref{tab:notation} gives the mathematical notation frequently used in the paper.

\begin{table}
    \caption{\scriptsize{Table of frequently used mathematical notation}}
    \label{tab:notation}
    \centering
    \begin{tabularx}{.5\textwidth}{l|l|l|X}
        \toprule
        \scriptsize{\textbf{Notation}}    & \scriptsize{\textbf{Description}} & \scriptsize{\textbf{Notation}}    & \scriptsize{\textbf{Description}} \\
        \midrule
        \scriptsize{$\naux$} & \scriptsize{\# auxiliary UAVs} & \scriptsize{$\bm{q}$} & \scriptsize{UAV configuration} \\
        \scriptsize{$\nobs$} & \scriptsize{\# obstacles} & {$\nw$} & \scriptsize{\# waypoints} \\
        \scriptsize{$\qobs$} & \scriptsize{Obstacle center position} &
        \scriptsize{$\Ts$} & \scriptsize{Sampling time} \\ 
        \scriptsize{$\bm{p}$} & \scriptsize{A point in 3D} & \scriptsize{$k$} & \scriptsize{time step counter} \\
        \bottomrule
    \end{tabularx}
\end{table}

\section{Background} \label{sec:background}

This paper focuses on controlling multi-UAV systems for wildfire suppression. 
These UAVs must coordinate multiple tracking tasks simultaneously. 
THC \cite{slotine_siciliano_1991} is a method that controls high-dimensional robots
to effectively optimize multiple tasks tracking objectives within an order of desired priorities. 
Using the null space projection method \cite{antonelli_et_al_2008}, the highest priority tasks are executed with the highest physical capabilities of the system, whereas tasks with lower priorities are executed using the remaining capabilities, avoiding interference with higher priority tasks. 
THC has been effectively applied to various robotics applications, including industrial manipulators \cite{liu_tan_padois_2016}
and UAV formations \cite{rosales_et_al_2016}. 
In our work, we utilize THC to effectively automate teams of UAVs that perform tasks needed for wildfire detection and suppression. 
Typical task tracking controllers are proportional-integral-derivative (PID) controllers.
Replacing them with PIC-based controllers, offers optimality beyond what PID controllers can achieve.
PIC \cite{kappen_2005} is a sampling-based stochastic optimal control method that optimizes an objective function that is formulated via a stochastic integral. 
PIC exploits the Monte-Carlo approach to compute the optimal control input using cost weighted average. 
PIC can be applied to nonlinear systems and in real-time using GPU resources. 
It has been applied in various robotics applications, e.g., for path following of a mobile robot \cite{zhu_guo_et_al_2019} 
and enabling a quadruped robot to jump or opening a door by a humanoid robot \cite{theodorou_et_al_2011}.
In this paper, a group of auxiliary UAVs monitors a fire extinguishing UAV and provides occlusion-free frames of view for the human operator that teleoperates it.
This is crucial for firefighting UAVs in searching the disaster scenes and in planning their actions \cite{queralta_et_al_2020}. 
Our proposed approach for determining these viewpoints is based on the model presented in \cite{dufek_et_al_2021}, where a team of 31 expert operators was asked to steer a robot to perform various navigation tasks 
from different viewpoints. 
The performance of the operators was scored,
and the highest-scored viewpoints were determined as the best viewpoints.
However, the original approach has not considered an environment with obstacles, that we introduce in our proposed control architecture.
Furthermore, we introduce a LoS obstacle avoidance method, that is different from approaches in literature because the LoS is employed to track a target rather than simply avoid obstacles. 
The current approaches include \cite{cichella_et_al_2018}, that only uses the LoS angle with respect to the obstacle instead of the obstacle position,
and \cite{wu_et_al_2006},
that perform obstacle avoidance based on
a reactive control strategy using fuzzy logic \cite{wu_et_al_2006},
using the LoS just for guidance purposes.
Our method, to the best of our knowledge, is the only one that uses the distance to the LoS to perform obstacle avoidance, while keeping the target inside the FoV of the sensor, therefore performing a more complex task while being collision-free.

\section{Problem statement} \label{sec:problem_statement}

The modeling and control paradigms of this paper are formulated for a 3D continuous space within a discrete time framework with sampling time $\Ts$ and step counter $k$. 
We consider one main UAV
and $\naux$ auxiliary UAVs, all equipped with a camera with angle $\theta$ for the conic FoV and with 
perfect perception. 
The main UAV, which is teleoperated by a human, flies over the fire to drop fire suppression materials on it. 
The auxiliary UAVs should autonomously monitor the main UAV and ensure that it is always within their FoV. 
The images captured by auxiliary UAVs provide the required information for the human operator to teleoperate the main UAV.
The configurations (i.e., the state including the 3D position) of the main and auxiliary UAVs are denoted, respectively, by $\qmain\in \mathbb{R}^{3}$ and $\bm{q}^\textup{aux}_u(k)\in \mathbb{R}^{3}$, for $u = 1, \ldots, \naux$. 
We consider $\nobs$ number of static, convex-shaped (spherical) obstacles in the environment, each
described by its center $\bm{q}_o \in \mathbb{R}^{3}$ and radius $r_o\in \mathbb{R}$.
The dynamics of each UAV is updated by:
\begin{align}
\label{eq:dynamics}
    \bm{x}(k + 1) = \bm{x}(k) + \bm{u}(k)\Ts 
\end{align}
Per time step, \eqref{eq:dynamics} is first used to update  
the velocity of the UAV under the impact of 
its acceleration, and then to 
update its configuration under the impact of its velocity.
We assume that each auxiliary UAV has perfect knowledge of its state and of the main UAV (using data captured 
by the cameras of all auxiliary UAVs).

\section{Proposed methodologies} \label{sec:methodologies}

Next, we elaborate our proposed approach and   
present a theorem that guarantees that the new LoS obstacle avoidance method continuously maintains a non-occluded view 
of the main UAV for the auxiliary UAVs.

\subsection{Autonomous control of auxiliary UAVs}
\label{sec:autonomous_control}

In fire suppression,
each auxiliary UAV $u$ 
follows
a trajectory determined by the PIC controller that steers the UAV towards its \textit{waypoints} 
(shown by $w_{u,i}$, for $u = 1, \ldots, \naux$ and $i= 1,\ldots,\nw_u$).  
The waypoints are determined based on the trajectory of the main UAV.
Per waypoint $w_{u,i}$, the autonomous control system performs the following $3$ actions:
\begin{enumerate}
    \item Determining the best viewpoints for the auxiliary UAVs
    \item Executing the tasks of the auxiliary UAVs  using THC
    \item Determining the location of the fire and the main UAV
\end{enumerate}
Next we explain how these actions are performed.

\subsubsection{Determining the best viewpoints}

To place themselves in the best viewpoints, the auxiliary UAVs should know the task that the main UAV 
is performing, including  
\textit{reachability}, i.e., flying above the fire, and \textit{manipulability}, i.e., dropping the suppression materials on fire. 
Based on the results of \cite{dufek_et_al_2021}, the best viewpoint locations 
for reachability and manipulability are, respectively, the front/top views and the right/top views with respect to the main UAV.
This is because the operator can better see the robot working from these views.

\subsubsection{Executing the task hierarchical control (THC)}

THC performs the next $4$ tasks
for the auxiliary UAVs:
\setlist[enumerate,1]{leftmargin=1.4cm}
\begin{enumerate}[label = \textbf{Task \arabic*}]
    \item 
    \label{task1}
    Avoiding collision with, e.g., trees or other UAVs
    \item 
    \label{task2}
    Maintaining a fixed distance with the main UAV
    \item 
    \label{task3}
    Avoiding obstacles on the line-of-sight (LoS), 
    i.e., a line that connects the auxiliary and main UAV positions, 
    to maintain an occlusion-free FoV 
    \item 
    \label{task4}
    Moving the UAV towards the best viewpoint
\end{enumerate}
\ref{task1} is activated whenever the auxiliary UAV is closer than a given threshold to an obstacle. 
\ref{task1} can be automated by defining a virtual local potential field 
of a sphere shape, denoted by   
$\mathcal{C}^{\text{obs}}$ (the center of which coincides with the center of the obstacle 
and its radius is the radius of the obstacle plus a safe distance), and by 
imposing a repulsive force on the auxiliary UAV in the radial direction \cite{qin_2017}. 
This force is obtained by differentiating the potential field function along the configuration 
vector of the auxiliary UAV. 
Maintaining a desired relative distance between an auxiliary and the main UAVs, i.e., \ref{task2}, 
is done in a similar way and is taken care of solely by the THC.
The THC sustains the distance
at (or around) a fixed value 
by defining a spherical (or ring-shaped) virtual local potential field, $\mathcal{C}^{\text{main}}$, centered at the main UAV.

\begin{figure} 
    \begin{center}
        \def\svgwidth{.7\columnwidth}
        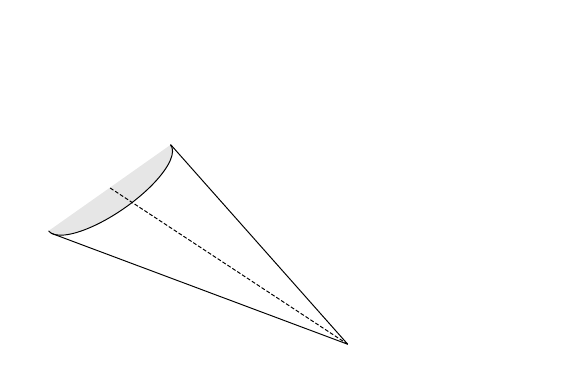
        \caption{\scriptsize{Illustration for avoiding obstacles on the LoS: $\pmain$ and $\paux$ are points where the main and auxiliary UAVs are respectively located; $\pobs$ is the center of volume of obstacle $\obs$; $\alpha$ is the imposed distance between the two UAVs; $\theta$ is the angle of the FoV; $\pone$, $\ptwo$ and $\pthree$ are intersection points. In the figure, we avoid the use of $k$ for making the illustrations less busy. }}
        \label{fig:theorem_scenario}
    \end{center}
\end{figure}

\ref{task3} guarantees that the LoS (i.e., the line segment that connects 
the auxiliary and main UAVs) is never occluded by the obstacle. 
In other words, a minimum safe distance $\gamma$ should be sustained between 
the conic FoV of the auxiliary UAV and the obstacle, 
thus between points $\ptwo$ and $\pthree$ (see Figure~\ref{fig:theorem_scenario}),  
i.e., the intersections of the line perpendicular to the LoS crossing through 
the center $\pobs$ of the obstacle with the surface of, respectively, the conic FoV of the auxiliary UAV and  
the obstacle. 
Similarly to the previous tasks, a spherical virtual potential field, 
$\mathcal{C}^{\text{LoS}}$, may be considered around the obstacle with a radius equal to the obstacle radius plus $\gamma$. A repulsive force must then be exerted on point $\ptwo$ in the radial direction. 
In reality, point $\ptwo$ is virtual and the auxiliary drone should be moved. 
This is done by translating the configuration of this virtual point after being repulsed, 
into the new configurations of the auxiliary UAV.%

With \ref{task4}, the auxiliary UAV moves to its next waypoint following the dynamics \eqref{eq:dynamics}.
The trajectories that the auxiliary UAVs follow in order to reach the waypoints are optimized using PIC, 
which determines the acceleration trajectory of the UAV. 
The waypoints for the auxiliary UAVs are determined based on the current task, i.e., reachability or manipulability, 
of the main UAV.
The time horizon of PIC is updated at every waypoint, and the PIC problem is resolved. 

\begin{figure*} 
    \centering
    \begin{tabular}{@{}c@{}}
        \def\svgwidth{0.25\textwidth}
        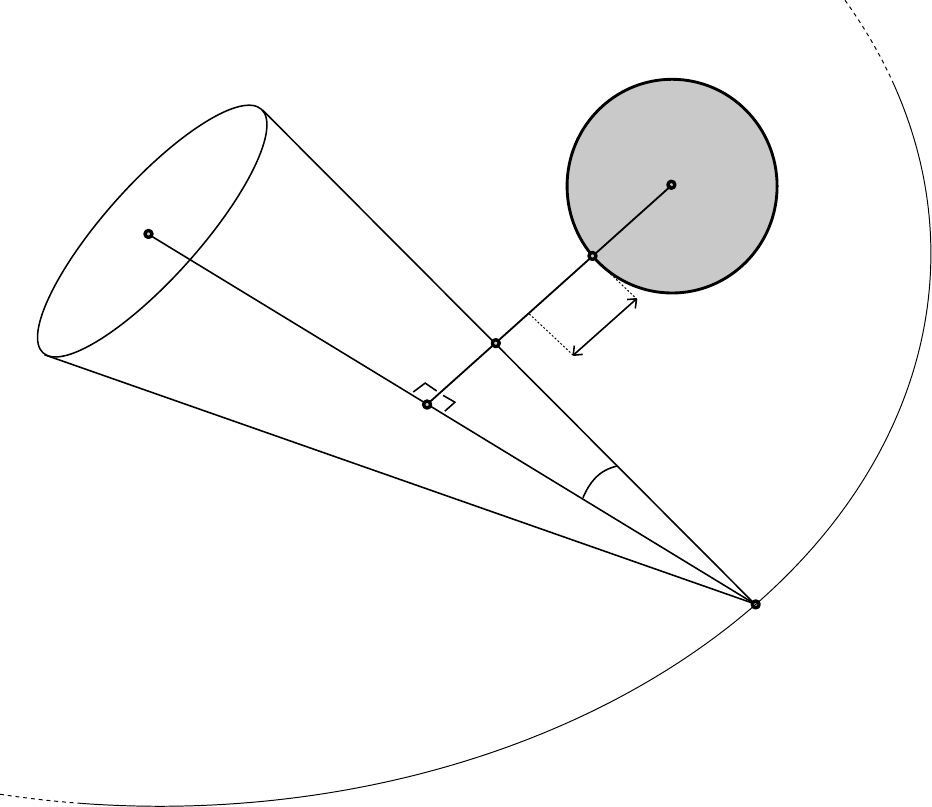
        \label{fig:theorem_case1}
        \hfil
    \end{tabular}
    \begin{tabular}{@{}c@{}}
        \def\svgwidth{0.25\textwidth}
        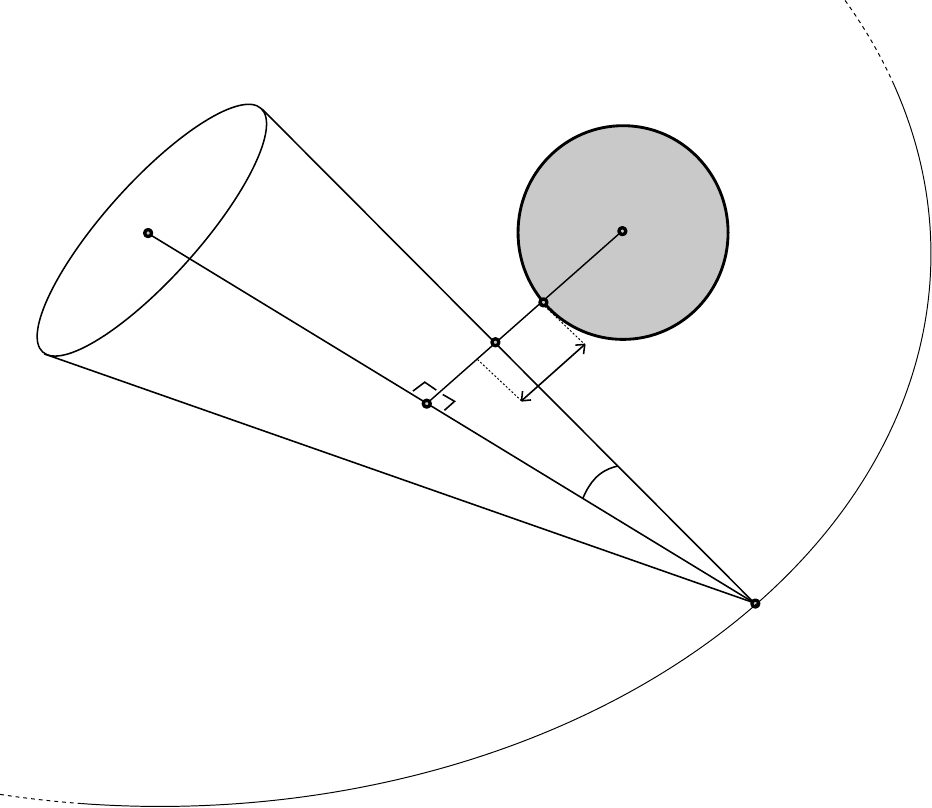
        \label{fig:theorem_case2}
    \end{tabular}
    \begin{tabular}{@{}c@{}}
        \def\svgwidth{0.25\textwidth}
        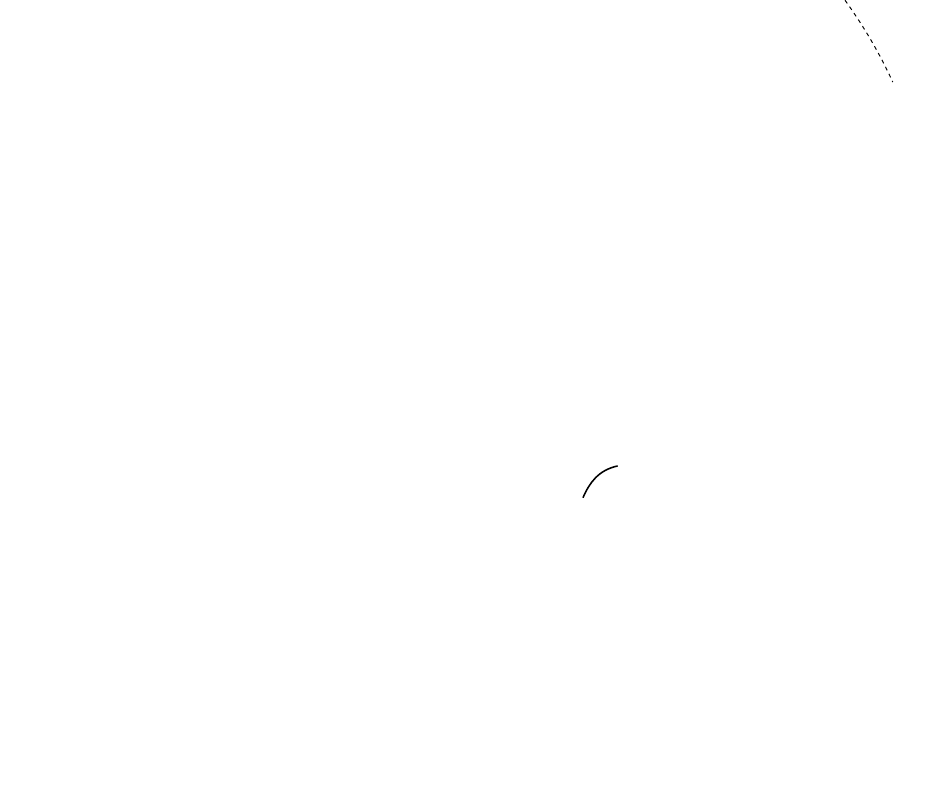
        \label{fig:theorem_case3}
    \end{tabular}
    \caption{\scriptsize{Cases of the threshold (1 to 3, respectively, from left to right): $\pmain$ and $\paux$ are the points where main UAV and auxiliary UAV are located, respectively, while $\pobs$ is the center of volume of obstacle $\obs$. Moreover, $\theta$ is the FoV angle, and $\pone$, $\ptwo$, $\pthree$ represent the considered intersection points. In the figure, we avoid the use of $k$ for making the illustrations less busy.}}
    \label{fig:theorem_cases}
\end{figure*}

\subsubsection{Determining the location of the fire and the main UAV}

The location of fire and the main UAV is determined via, e.g., computer vision techniques applied to 
the images that are captured by the auxiliary UAVs. 
In this paper, we assume that these locations are already known to the auxiliary UAVs.

\subsection{Theorem: LoS obstacle avoidance}
\label{sec:theorem}

We next present a theorem to show that, with the proposed methods, 
the auxiliary UAVs will always have the main UAV inside their FoVs, which will always be occlusion-free.
Although we consider one auxiliary UAV, the theorem is generalizable to multi-UAV systems, 
since Task~1 of THC ensures that all auxiliary UAVs are far enough from each other when LoS obstacle avoidance 
is activated through Task~3. Hence, the required distance between each auxiliary UAV and the obstacle is 
maintained independently of other auxiliary UAVs. 
We define $\qaux, \qmain \in \mathbb{R}^{3}$ as the configuration of, respectively, 
the auxiliary and the main UAV, $\paux$ and $\pmain$ as their representing points in the 3D space, 
and $\pauxpmain$ (where $\overline{\bm{p}_\textrm{a} \bm{p}_\textrm{b}}$ indicates the line segment between points $\bm{p}_\textrm{a}$ and $\bm{p}_\textrm{b}$) the LoS, which is also the principal axis of the cone 
with vertex $\paux$ that represents the FoV of the auxiliary UAV when oriented towards the main UAV (see Figure~\ref{fig:theorem_scenario}). 
Moreover, $\pobs$ is the center of the spherical obstacle $\obs$, and $\Vobs$ and $\Vfov$ 
are the (continuous) sets 
that include all points on the outer surface and inside volume of, respectively, the obstacle 
and the FoV of the auxiliary UAV.

\begin{theorem} \label{theorem_LoS_obst_avoid}
    By including Task~3, ``avoiding obstacles on the LoS'', in the THC framework, the FoV of the auxiliary UAV always remains occlusion-free (i.e., $\Vfov \cap \Vobs = \emptyset$), and a distance larger than or equal to a specified threshold $\gamma$ is maintained between the FoV and the obstacle, i.e., $\lVert \qthree - \qtwo \rVert > \gamma$ 
    (see $\ptwo \in \Vfov$ and $\pthree \in \Vobs$ in Figure~\ref{fig:theorem_scenario}).
\end{theorem}

\begin{figure*} [t]
    \begin{center}
        \def\svgwidth{0.7\textwidth}
\begingroup%
  \makeatletter%
  \providecommand\color[2][]{%
    \errmessage{(Inkscape) Color is used for the text in Inkscape, but the package 'color.sty' is not loaded}%
    \renewcommand\color[2][]{}%
  }%
  \providecommand\transparent[1]{%
    \errmessage{(Inkscape) Transparency is used (non-zero) for the text in Inkscape, but the package 'transparent.sty' is not loaded}%
    \renewcommand\transparent[1]{}%
  }%
  \providecommand\rotatebox[2]{#2}%
  \newcommand*\fsize{\dimexpr\f@size pt\relax}%
  \newcommand*\lineheight[1]{\fontsize{\fsize}{#1\fsize}\selectfont}%
  \ifx\svgwidth\undefined%
    \setlength{\unitlength}{1355.25bp}%
    \ifx\svgscale\undefined%
      \relax%
    \else%
      \setlength{\unitlength}{\unitlength * \real{\svgscale}}%
    \fi%
  \else%
    \setlength{\unitlength}{\svgwidth}%
  \fi%
  \global\let\svgwidth\undefined%
  \global\let\svgscale\undefined%
  \makeatother%
  \begin{picture}(1,0.52185944)%
    \lineheight{1}%
    \setlength\tabcolsep{0pt}%
    \put(0,0){\includegraphics[width=\unitlength,page=1]{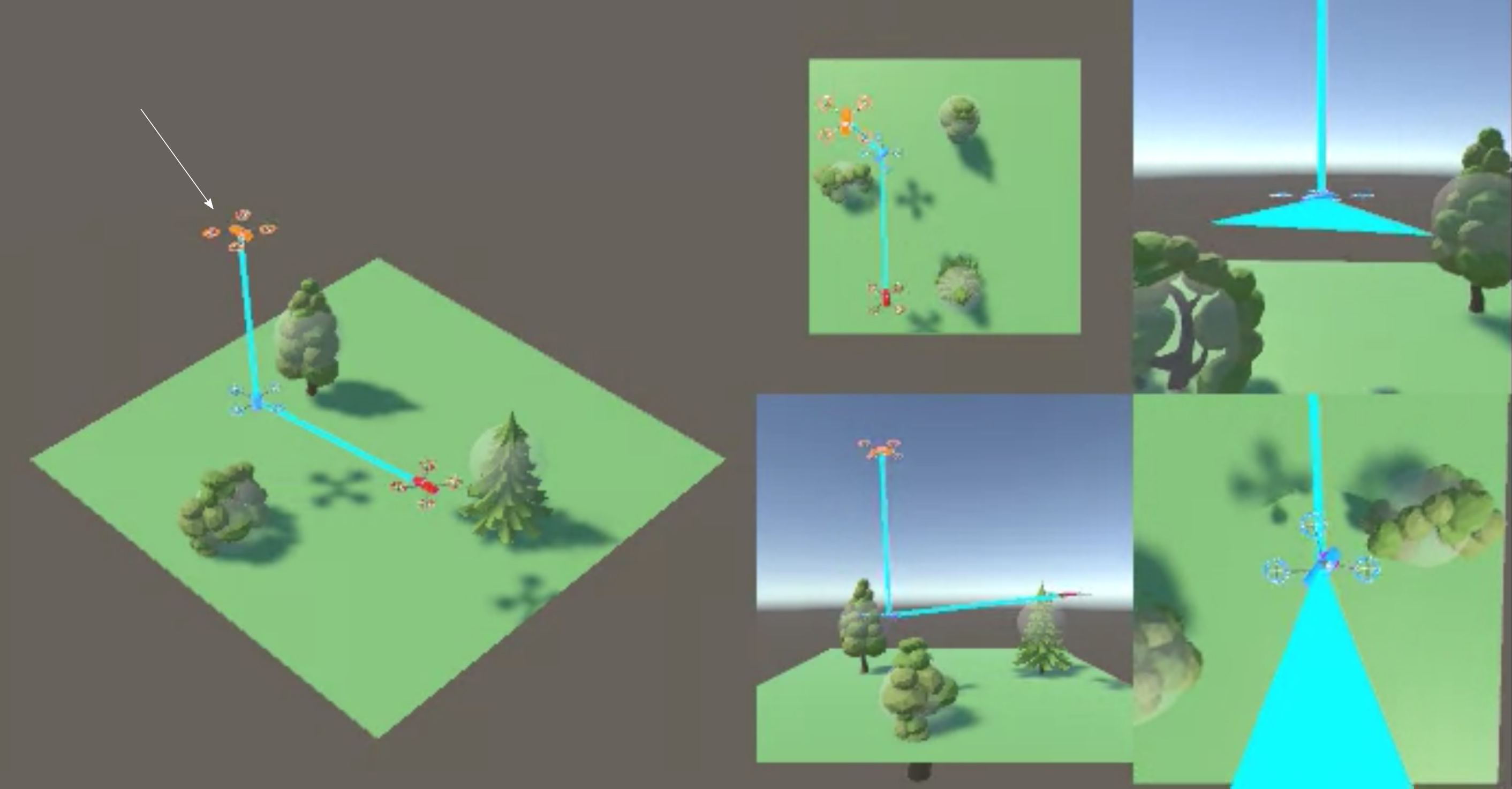}}%
    \put(0.04500000,0.45719271){\makebox(0,0)[lt]{\lineheight{1.25}\smash{\begin{tabular}[t]{l}{\footnotesize \color{white} Auxiliary UAV 2} \end{tabular}}}}%
    \put(0,0){\includegraphics[width=\unitlength,page=2]{unity_simulations_overview.pdf}}%
     \put(0.04286093,0.32183497){\makebox(0,0)[lt]{\lineheight{1.25}\smash{\begin{tabular}[t]{l}{\footnotesize \color{white} Main UAV} \end{tabular}}}}%
   \put(0,0){\includegraphics[width=\unitlength,page=3]{unity_simulations_overview.pdf}}%
    \put(0.30964054,0.33114062){\makebox(0,0)[lt]{\lineheight{1.25}\smash{\begin{tabular}[t]{l}{\footnotesize \color{white} Auxiliary UAV 1} \end{tabular}}}}%
    \put(0,0){\includegraphics[width=\unitlength,page=4]{unity_simulations_overview.pdf}}%
  \end{picture}%
\endgroup%

        \caption{\scriptsize{Illustration of the scenario simulated in Unity: From left to right, the 3D view of the simulated outdoor environment, the top and side views of the environment, and views of the cameras of the two auxiliary UAVs. The main UAV has cyan color, the auxiliary UAVs have red and orange colors.}}
        \label{fig:unity_simulations_overview}
    \end{center}
\end{figure*}

\noindent

\begin{proof}
    Let $\pone$ be the closest point of $\pauxpmain$ to $\pobs$, 
    $\ponepobs$ be a line segment perpendicular to $\pauxpmain$,  
    and $\ptwo \in \Vfov$ and $\pthree \in \Vobs$ be the closest points to each other on $\ponepobs$ (see Figure~\ref{fig:theorem_scenario}). 
    Since obstacle $\obs$ is convex-shaped, its intersection point $\pthree$ with $\ponepobs$ is unique.
    Consider $\qone, \qtwo, \qthree$ as the configurations of points $\pone, \ptwo, \pthree$, respectively. 
    Supposing a specified threshold value $\gamma > 0$, one of the following $3$ cases will occur (see  Figure~\ref{fig:theorem_cases}):  
    \setlist[enumerate,1]{leftmargin=1.5cm}
    \begin{enumerate}[start=0,label={Case \arabic*}]
        \setcounter{enumi}{0}
        \item 
        \label{case1}
        $\lVert \qtwo - \qone \rVert + \gamma < \lVert \qthree - \qone \rVert$
        \item 
        \label{case2}
        $\lVert \qtwo - \qone \rVert < \lVert \qthree - \qone \rVert$ \& $\lVert \qtwo - \qone \rVert + \gamma > \lVert \qthree - \qone \rVert$
        \item 
        \label{case3}
        $\lVert \qtwo - \qone \rVert > \lVert \qthree - \qone \rVert$
    \end{enumerate}
    For \ref{case1}, evidently $\Vfov \cap \Vobs = \emptyset$ and $\lVert \qthree - \qtwo \rVert > \gamma$  
    and, thus, Theorem~\ref{theorem_LoS_obst_avoid} holds.
    For \ref{case2} and \ref{case3}, however, the situation is different:  
    In \ref{case2}, although $\lVert \qtwo - \qone \rVert < \lVert \qthree - \qone \rVert$ implies $\Vfov \cap \Vobs = \emptyset$, we have $\lVert \qthree - \qtwo \rVert < \gamma$ (see the middle plot in Figure~\ref{fig:theorem_cases}). 
    In \ref{case3}, none of the two conditions holds, i.e., $\lVert \qtwo - \qone \rVert > \lVert \qthree - \qone \rVert$ implies $\Vfov \cap \Vobs \neq \emptyset$ and  $\ptwo \in \Vobs$ (see the third plot in Figure~\ref{fig:theorem_cases}). 
    In these cases \ref{task3} will be activated, and due 
    to the prioritized performance of the tasks via THC,
    this implies that both \ref{task1} and \ref{task2} have already been activated.
    Thus, all corresponding scalar potential fields must be integrated through a (weighted) summation. 
    Let us show the combined virtual potential field function by $\Phi(\qmain, \qaux, \qobs)$ and  
    the task error by:
    \begin{equation}
        \label{eq:error}
        \bm{e}_3(k) = \lVert \qthree - \qtwo \rVert - \gamma
    \end{equation}
    By defining a candidate Lyapunov function $V(k) = \frac{1}{2}\bm{e}^{\top}_3(k) k_1 \bm{e}_3(k) + k_2 \Phi(\qmain, \qaux, \qobs)$, 
    one can show that
    the time derivative of this Lyapunov function and thus, 
    the task error \eqref{eq:error} will converge to zero, when using the following control input:
    \begin{equation}
        \label{eq:control_input_part1}
        \bm{u}_3(k) = J^+(\Phi(\qmain, \qaux, \qobs)) \biggl( k^\mathrm{p} \bm{e}_3(k) \nonumber
    \end{equation}
    \begin{equation}
        \label{eq:control_input_part2}
        + k^\mathrm{d} \frac{\mathrm{d} \bm{e}_3(k)}{\mathrm{d} t} + k^\mathrm{i} \int \bm{e}_3(k) \mathrm{d} t \biggl)
    \end{equation}
    with $J^+$ the pseudo-inverse of the Jacobian function, and $k^\mathrm{p}$, $k^\mathrm{d}$ and $k^\mathrm{i}$ gains that must be tuned.
\end{proof}

\begin{remark}    
    Note that Theorem~\ref{theorem_LoS_obst_avoid} holds for one obstacle. 
    To extend it to more obstacles, one needs to ensure that the exclusion of the admissible set 
    of the states of the auxiliary UAV and the points that 
    construct the part of the world that is covered by the virtual potential field $\Phi(\qmain, \qaux, \qobs)$ 
    is non-empty. Whenever this condition does not hold, the LoS obstacle avoidance problem will be infeasible.
\end{remark}
\begin{remark} \label{remark_convex_hull}
    In case of a non-convex obstacle, the same THC approach may be used but to guarantee 
    Theorem~\ref{theorem_LoS_obst_avoid}, the obstacle may be represented by its convex hull.
\end{remark}

\section{Simulations and results} \label{sec:simulations}

This section explains the case study that was performed for evaluating the proposed approaches and presents the results.

\subsection{Setup}

All algorithms, including the THC controller, used in this case study were implemented in MATLAB (R2021a version) 
and the 3D simulations of the UAVs in outdoor missions were developed in Unity (2021 version). 
An overview of the simulated environments is illustrated in Figure~\ref{fig:unity_simulations_overview}. 
The PC, on which the experiments were conducted, had Intel Core i7 processor with 4 cores at 1.80 GHz-2.30 GHz and 16 GB RAM.
We consider $2$ auxiliary UAVs and a cubic-shaped environment of size $22 \times 22 \times 22 \; \mathrm{m}^3$, in which $3$ obstacles are present.
The comparisons were performed in 5 different scenarios: (1) Trees present as obstacles. 
(2) Flying obstacles present (e.g., other UAVs). (3) Dangerous ground obstacles present (e.g., trees that are 
taller, thus closer to the UAVs, compared to scenario~(1)); (4) The main UAV changing its task during the scenario 
affecting the position of the auxiliary UAVs. (5) The main UAV moves across an $8$-shaped trajectory that is challenging for UAVs to follow.

\subsection{Comparison and results} \label{sec:comparison_results}

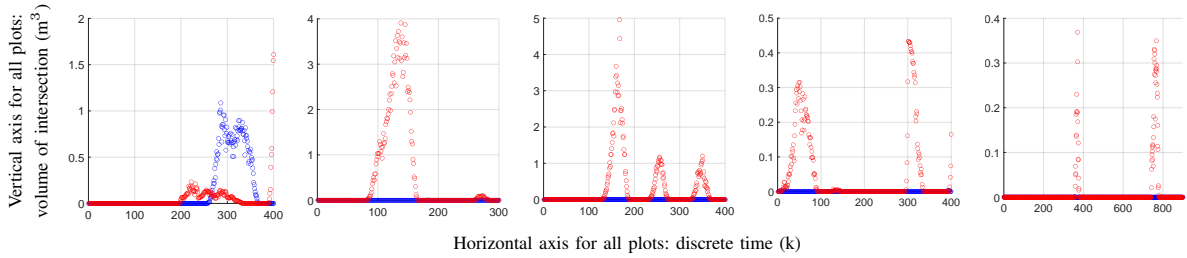
\begin{figure*}
    \centering
    \begin{tikzpicture}[
        /pgfplots/every axis/.style={
            width=5cm,
            height=5cm,
            at={(0,0)},
            ymin=-5,
            ymax=5,
            xmin=0,
            xmax=5, 
            no marks, 
            ylabel=Intensity (a.u.)                
        },
        /pgfplots/table/col sep=comma,
        /pgfplots/table/x=time,
        /pgfplots/table/y=real,
        label box/.style={
            anchor=north east,
            at={(axis cs:5,5)},
            fill=white,
            inner sep=6pt,
            font=\sffamily\bfseries\Large,
        },
        shared axis box/.style={
            outer sep=0.1pt,
            inner sep=0.1pt,
            minimum width=0.1cm,
            anchor=north,
        }
    ]
        \begin{axis}[
            axis lines=none,
            xtick=\empty, 
            ytick=\empty,
        ]
            \node[label box] {\includegraphics[width=.16\textwidth]{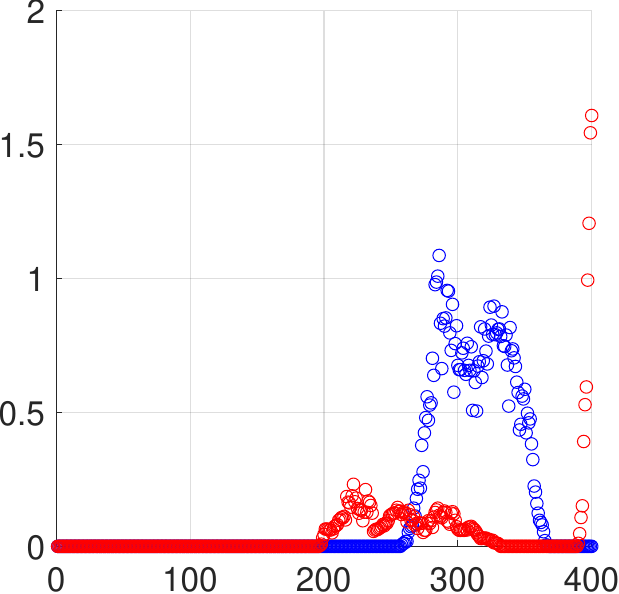}};
        \end{axis}
        \begin{axis}[
                xshift=3cm,
                ylabel={},
                thick,
                yticklabel=\empty,
                xticklabel=\empty,
                axis lines=none, 
                xtick=\empty, 
                ytick=\empty
            ]
            \node[label box] {\includegraphics[width=.15\textwidth]{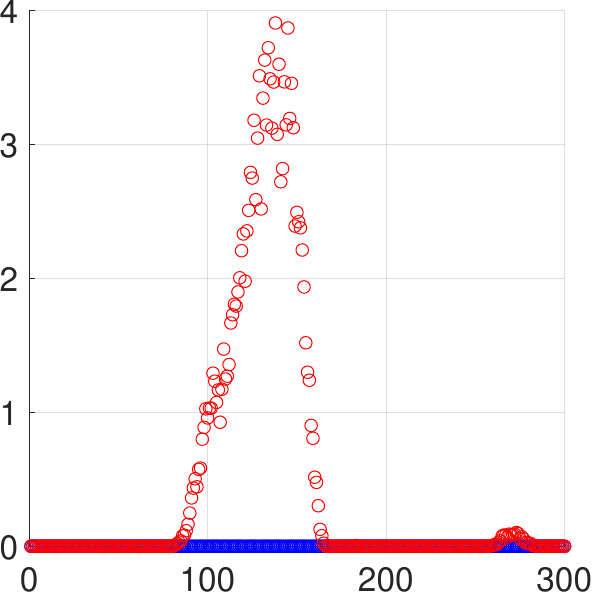}};
        \end{axis}
        \begin{axis}[
            xshift=6cm,
            axis lines=none,
            xtick=\empty, 
            ytick=\empty,
        ]
            \node[label box] {\includegraphics[width=.15\textwidth]{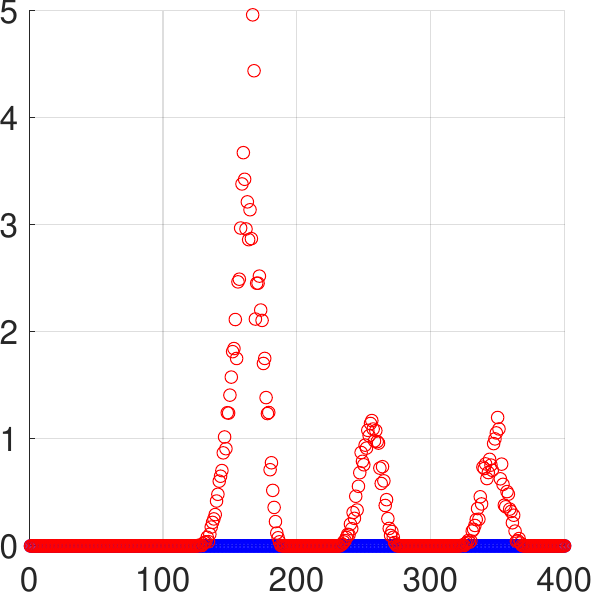}};
        \end{axis}
        \begin{axis}[
            xshift=9cm,
            axis lines=none,
            xtick=\empty, 
            ytick=\empty,
        ]
            \node[label box] {\includegraphics[width=.15\textwidth]{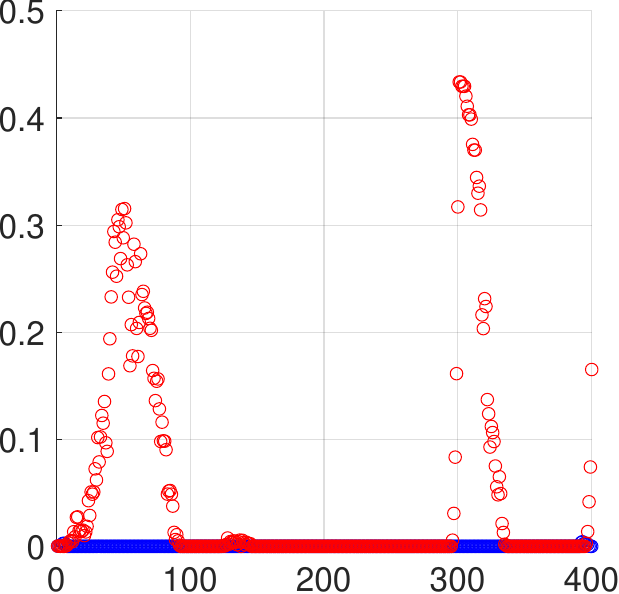}};
        \end{axis}
        \begin{axis}[
            xshift=12cm,
            axis lines=none,
            xtick=\empty, 
            ytick=\empty,
        ]
            \node[label box] {\includegraphics[width=.15\textwidth]{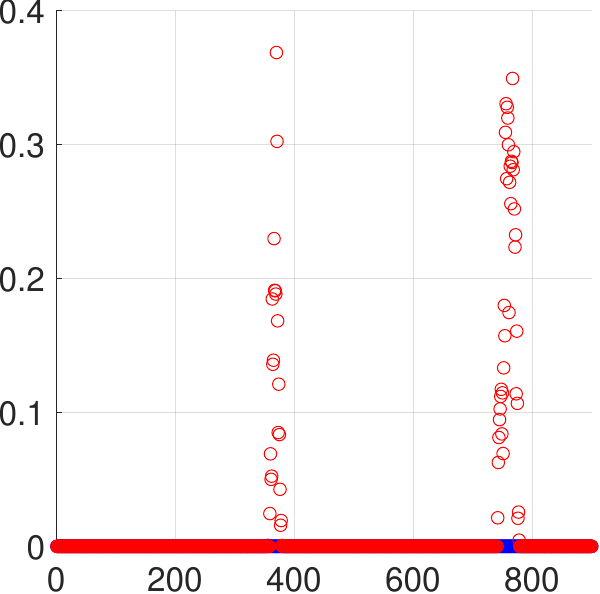}};
        \end{axis}
    
    \node[shared axis box] at (current bounding box.south) {\scriptsize Horizontal axis for all plots: discrete time (k)};

    \node[shared axis box, xshift=-.3cm, yshift=0.2cm, rotate=90] at (current bounding box.west) {\scriptsize volume of intersection ($\text{m}^3$)};

    \node[shared axis box, xshift=-.3cm, yshift=0.2cm, rotate=90] at (current bounding box.west) {\scriptsize Vertical axis for all plots:};

    \end{tikzpicture}
    
    \caption{\scriptsize{Comparison of the intersection of the FoV with obstacles, with LoS obstacle avoidance task active (blue) or inactive (red): volume of intersection of FoV with obstacles (occluded view). Plots from left to right represent the results for scenarios from 1 to 5.}}
    
    \label{fig:fov_intersection_volume}

\end{figure*}

\begin{figure*}
    \centering
    \begin{tikzpicture}[
        /pgfplots/every axis/.style={
            width=5cm,
            height=5cm,
            at={(0,0)},
            ymin=-5,
            ymax=5,
            xmin=0,
            xmax=5, 
            no marks, 
            ylabel=Intensity (a.u.)                
        },
        /pgfplots/table/col sep=comma,
        /pgfplots/table/x=time,
        /pgfplots/table/y=real,
        label box/.style={
            anchor=north east,
            at={(axis cs:5,5)},
            fill=white,
            inner sep=6pt,
            font=\sffamily\bfseries\Large,
        },
        shared axis box/.style={
            outer sep=0.11pt,
            inner sep=0.11pt,
            minimum width=0.1cm,
            anchor=north,
        }
    ]
        \begin{axis}[
            axis lines=none,
            xtick=\empty, 
            ytick=\empty,
        ]
            \node[label box] {\includegraphics[width=.15\textwidth]{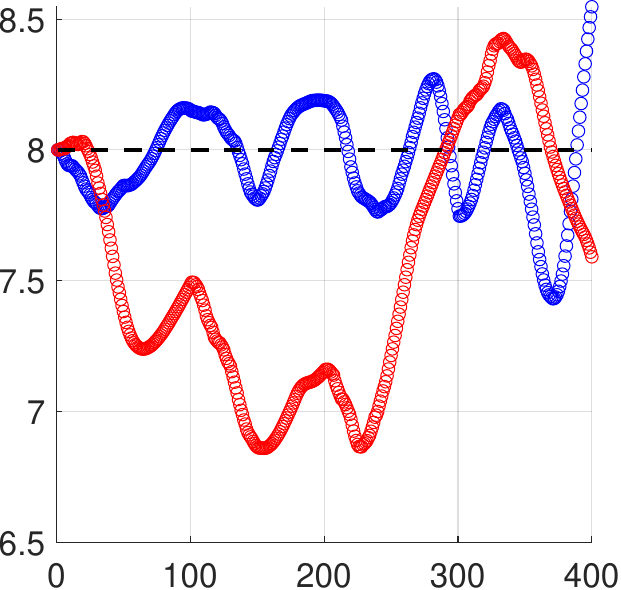}};
        \end{axis}
        \begin{axis}[
                xshift=3cm,
                ylabel={},
                thick,
                yticklabel=\empty,
                xticklabel=\empty,
                axis lines=none, 
                xtick=\empty, 
                ytick=\empty
            ]
            \node[label box] {\includegraphics[width=.15\textwidth]{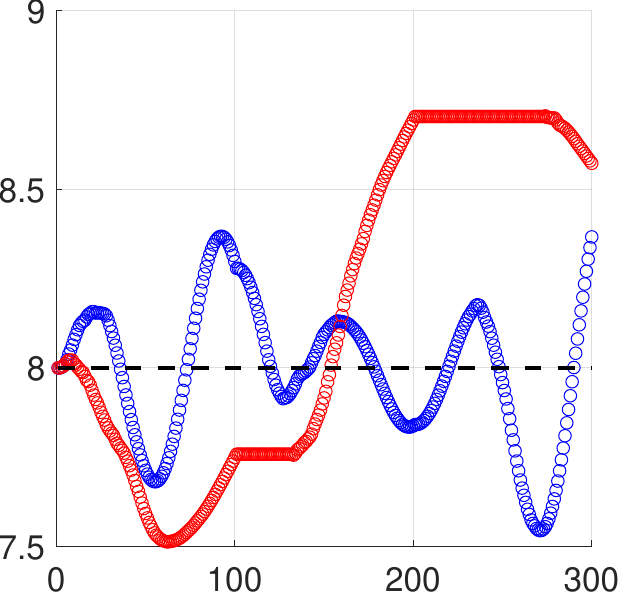}};
        \end{axis}
        \begin{axis}[
            xshift=6cm,
            axis lines=none,
            xtick=\empty, 
            ytick=\empty,
        ]
            \node[label box] {\includegraphics[width=.15\textwidth]{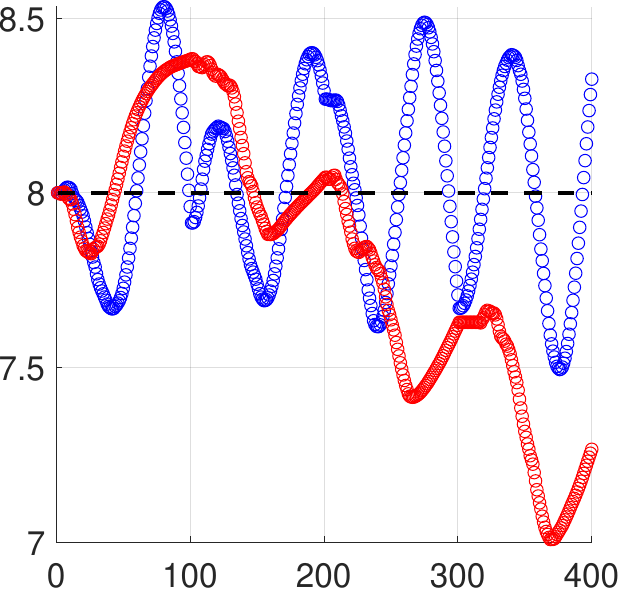}};
        \end{axis}
        \begin{axis}[
            xshift=9cm,
            axis lines=none,
            xtick=\empty, 
            ytick=\empty,
        ]
            \node[label box] {\includegraphics[width=.15\textwidth]{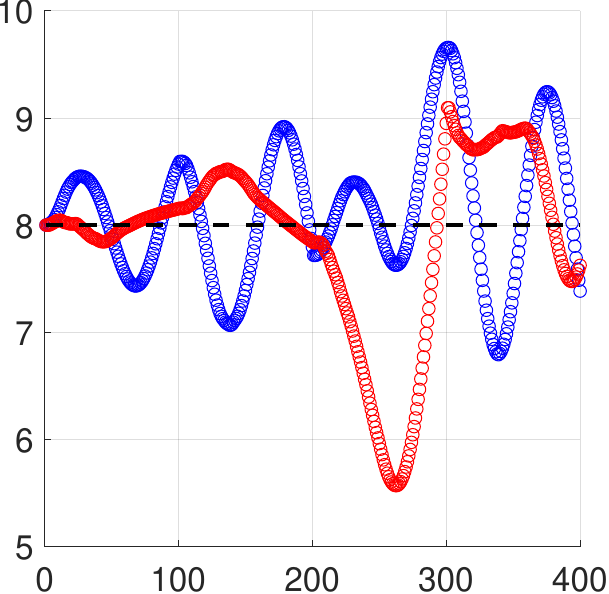}};
        \end{axis}
        \begin{axis}[
            xshift=12cm,
            axis lines=none,
            xtick=\empty, 
            ytick=\empty,
        ]
            \node[label box] {\includegraphics[width=.15\textwidth]{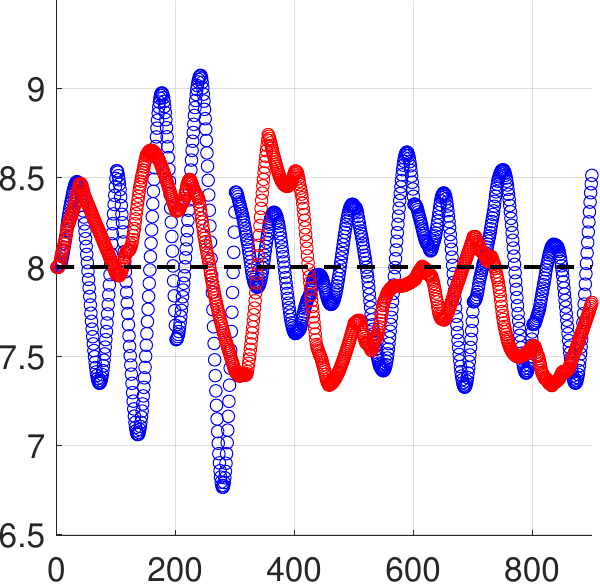}};
        \end{axis}

        \begin{axis}[
            yshift=-3.5cm,
            axis lines=none,
            xtick=\empty, 
            ytick=\empty,
        ]
            \node[label box] {\includegraphics[width=.15\textwidth]{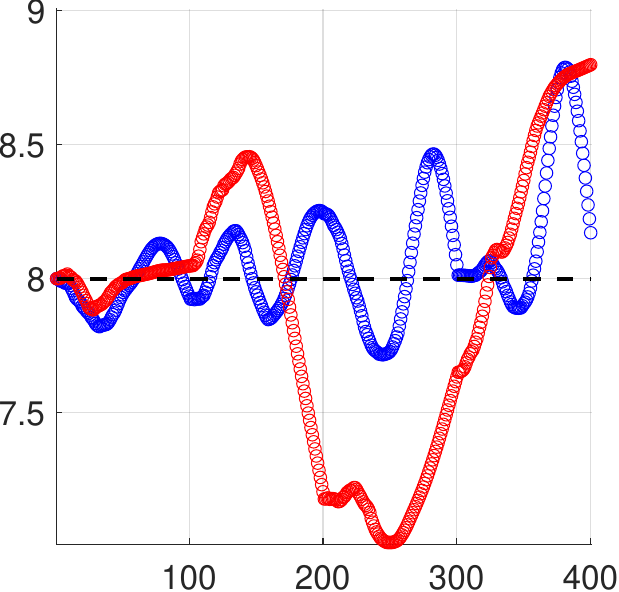}};
        \end{axis}
        \begin{axis}[
                xshift=3cm,
                yshift=-3.5cm,
                ylabel={},
                thick,
                yticklabel=\empty,
                xticklabel=\empty,
                axis lines=none, 
                xtick=\empty, 
                ytick=\empty
            ]
            \node[label box] {\includegraphics[width=.15\textwidth]{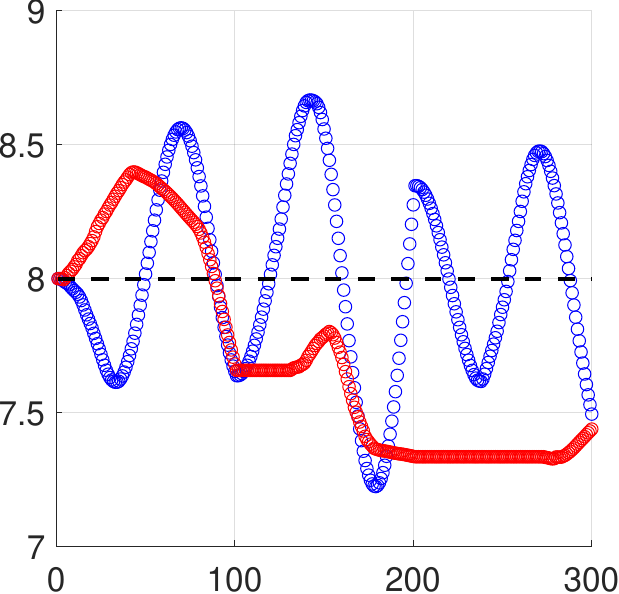}};
        \end{axis}
        \begin{axis}[
            xshift=6cm,
            yshift=-3.5cm,
            axis lines=none,
            xtick=\empty, 
            ytick=\empty,
        ]
            \node[label box] {\includegraphics[width=.15\textwidth]{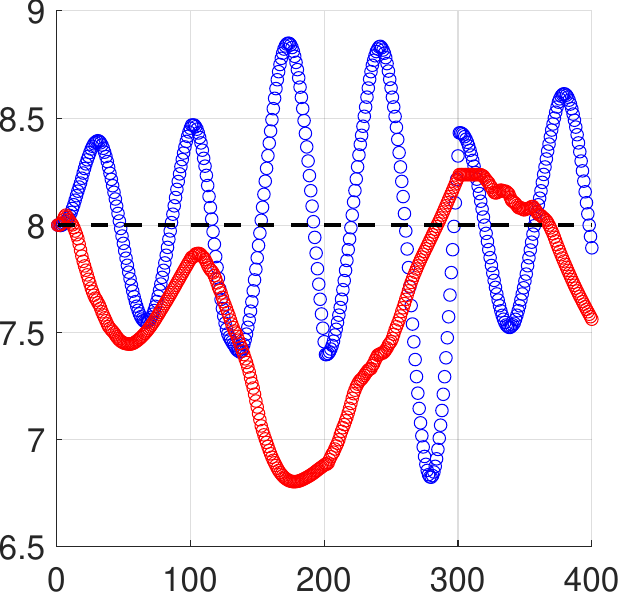}};
        \end{axis}
        \begin{axis}[
            xshift=9cm,
            yshift=-3.5cm,
            axis lines=none,
            xtick=\empty, 
            ytick=\empty,
        ]
            \node[label box] {\includegraphics[width=.15\textwidth]{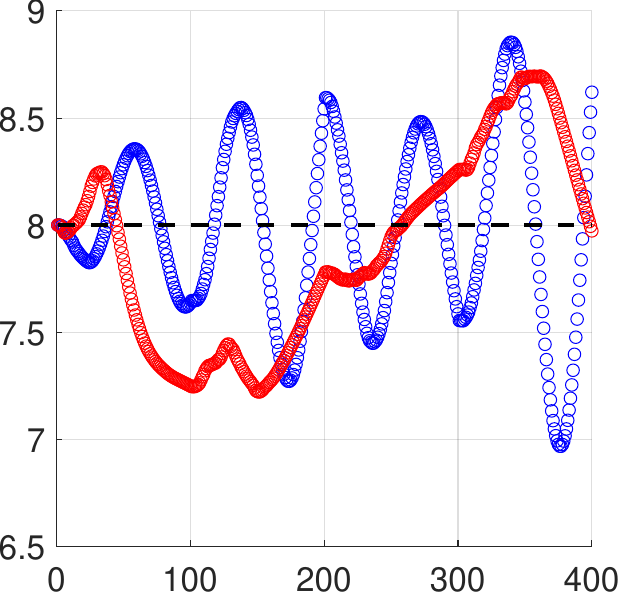}};
        \end{axis}
        \begin{axis}[
            xshift=12cm,
            yshift=-3.5cm,
            axis lines=none,
            xtick=\empty, 
            ytick=\empty,
        ]
            \node[label box] {\includegraphics[width=.15\textwidth]{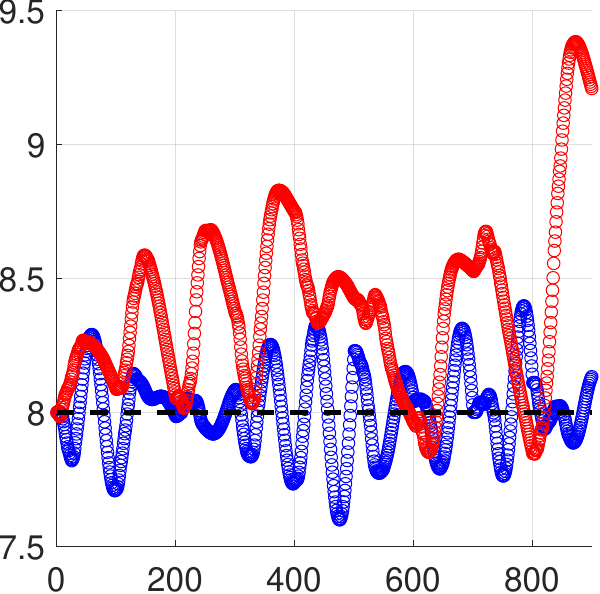}};
        \end{axis}
    
    \node[shared axis box] at (current bounding box.south) {\scriptsize Horizontal axis for all plots: discrete time (k)};

    \node[shared axis box, xshift=-.3cm, yshift=0.2cm, rotate=90] at (current bounding box.west) {\scriptsize distance between aux. UAV and main UAV (m)};

    \node[shared axis box, xshift=-.4cm, yshift=0.2cm, rotate=90] at (current bounding box.west) {\scriptsize Vertical axis for all plots:};

    \end{tikzpicture}
    
    \caption{\scriptsize{Comparison of the distance between 
    main UAV and
    auxiliary UAV 1 (top plots)
    and 2 (bottom plots),
    with the distance task active (blue curves) or inactive (red curves). The black dashed lines indicate the desired distance. Plots from left to right represent the results for scenarios from 1 to 5.}}
    
    \label{fig:robots_distance}

\end{figure*}
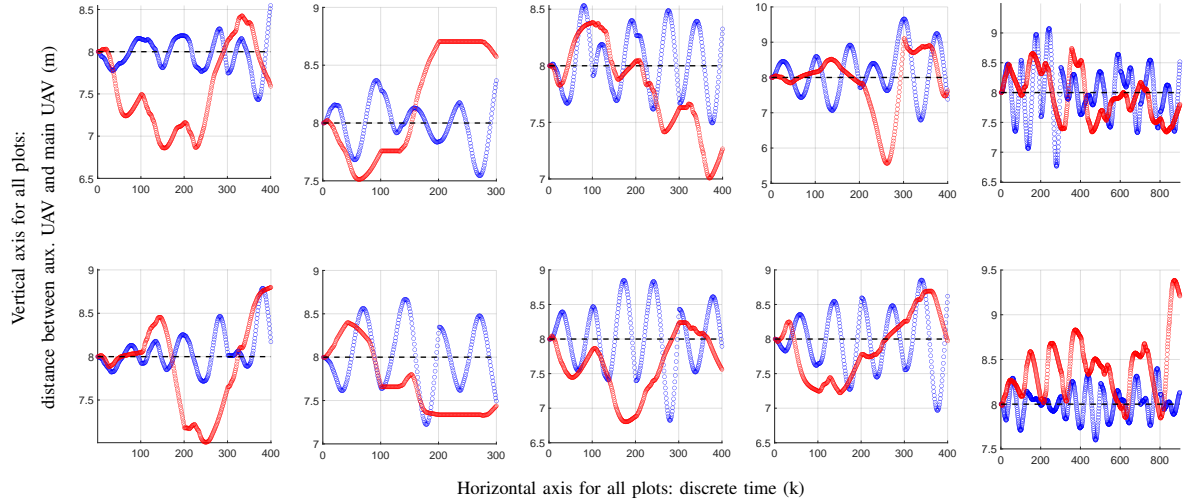

We performed an ablation study, in which we compared the case where some components of the THC controller were excluded and the results were compared with the complete framework of THC being implemented:
with and without LoS obstacle avoidance task, with and without distance maintenance task, and with and without PIC in the target-reaching task, for each scenario.
For the LoS obstacle avoidance task active or inactive in THC, in Table~\ref{tab:fov_intersection_time_LoS_comparison} we show the percentage of time
in which the FoVs of the auxiliary UAVs intersect any obstacles (i.e., occluded view),
and in Figure~\ref{fig:fov_intersection_volume} we show the volume of intersection of the FoVs with any obstacles,
with respect to time. 
Then, for the distance task active or inactive, 
in Figure~\ref{fig:robots_distance} we show the distance between the main UAV and each auxiliary UAV,
with respect to time,
and in Table~\ref{tab:distance_errors} we show the maximum and average errors, respectively, with respect to the assigned viewpoint distance (between main and auxiliary UAVs), for each auxiliary UAV.
Finally, with PIC or without PIC (i.e., using PID controllers), 
in Table~\ref{tab:PIC_comparison} we show the length of the path of each UAV,
to generate the trajectory to the goal,
and the percentage of time in which the FoV is intersecting any obstacle, to analyze the performance in maintaining the sight.

\begin{table}[t]
    \caption{\scriptsize{Percentage of time that the intersection of FoV and obstacles is non-empty:
    The bold values indicate smaller intersections, thus less occlusion.}}
    \label{tab:fov_intersection_time_LoS_comparison}
    \centering
    \begin{tabular}{|c||c|c|}
        \hline
        & \scriptsize{Task~3 inactive} & \scriptsize{Task~3 active} \\
        \hline \hline
        \scriptsize{Scenario 1} & \scriptsize{40.25\%} & \scriptsize{\textbf{30.25\%}} \\
        \hline
        \scriptsize{Scenario 2} & \scriptsize{44.00\%} & \scriptsize{\textbf{0.00}} \\
        \hline
        \scriptsize{Scenario 3} & \scriptsize{41.50\%} & \scriptsize{\textbf{1.50\%}} \\
        \hline
        \scriptsize{Scenario 4} & \scriptsize{44.75\%} & \scriptsize{\textbf{12.75\%}} \\
        \hline
        \scriptsize{Scenario 5} & \scriptsize{8.67\%} & \scriptsize{\textbf{6.33\%}} \\
        \hline
    \end{tabular}
\end{table}

\begin{table}[t]
    \caption{\scriptsize{Maximum error and average error with respect to the assigned viewpoint distance, with \ref{task2}
    inactive or active. The bold values indicate smaller errors.}}
    \label{tab:distance_errors}
    \centering
    \begin{tabular}{|c||c|c|c|c|}
        \hline
        \multirow{2}{*}{} & \multicolumn{2}{c|}{\scriptsize{Maximum error}}& \multicolumn{2}{c|}{\scriptsize{Average error}} \\
        \cline{2-5}
        & \scriptsize{Inactive} & \scriptsize{Active} & \scriptsize{Inactive} & \scriptsize{Active} \\
        \hline \hline
        \scriptsize{Scenario 1: aux. 1 UAV} & \scriptsize{1.14 m} & \scriptsize{\textbf{0.57 m}} & \scriptsize{0.55 m} & \scriptsize{\textbf{0.16 m}} \\
        \hline
        \scriptsize{Scenario 1: aux. 2 UAV} & \scriptsize{0.99 m} & \scriptsize{\textbf{0.79 m}} & \scriptsize{0.40 m} & \scriptsize{\textbf{0.17 m}} \\
        \hline
        \scriptsize{Scenario 2: aux. 1 UAV} & \scriptsize{0.71 m} & \scriptsize{\textbf{0.46 m}} & \scriptsize{0.42 m} & \scriptsize{\textbf{0.16 m}} \\
        \hline
        \scriptsize{Scenario 2: aux. 2 UAV} & \scriptsize{\textbf{0.67 m}} & \scriptsize{0.78 m} & \scriptsize{0.43 m} & \scriptsize{\textbf{0.31 m}} \\
        \hline
        \scriptsize{Scenario 3: aux. 1 UAV} & \scriptsize{0.99 m} & \scriptsize{\textbf{0.53 m}} & \scriptsize{0.33 m} & \scriptsize{\textbf{0.23 m}} \\
        \hline
        \scriptsize{Scenario 3: aux. 2 UAV} & \scriptsize{1.20 m} & \scriptsize{\textbf{1.18 m}} & \scriptsize{0.44 m} & \scriptsize{\textbf{0.39 m}} \\
        \hline
        \scriptsize{Scenario 4: aux. 1 UAV} & \scriptsize{2.43 m} & \scriptsize{\textbf{1.66 m}} & \scriptsize{0.56 m} & \scriptsize{\textbf{0.53 m}} \\
        \hline
        \scriptsize{Scenario 4: aux. 2 UAV} & \scriptsize{\textbf{0.78 m}} & \scriptsize{1.03 m} & \scriptsize{0.40 m} & \scriptsize{\textbf{0.35 m}} \\
        \hline
        \scriptsize{Scenario 5: aux. 1 UAV} & \scriptsize{\textbf{0.74 m}} & \scriptsize{1.23 m} & \scriptsize{\textbf{0.34 m}} & \scriptsize{\textbf{0.34 m}} \\
        \hline
        \scriptsize{Scenario 5: aux. 2 UAV} & \scriptsize{1.38 m} & \scriptsize{\textbf{0.40 m}} & \scriptsize{0.41 m} & \scriptsize{\textbf{0.12 m}} \\
        \hline
    \end{tabular}
\end{table}

\begin{table}[t]
    \caption{\scriptsize{Path length of each UAV and percentage of time that with non-empty intersection of FoV and obstacles, with \ref{task4}
    performed using PID or PIC. The bold values indicate, respectively, shorter paths and smaller intersections.}}
    \label{tab:PIC_comparison}
    \centering
    \begin{tabular}{|c|c||c|c|c|c|}
        \hline
        \multirow{2}*{} & & \multicolumn{2}{c|}{\scriptsize{Path length}} & \multicolumn{2}{c|}{\scriptsize{Intersection time}} \\
        \cline{2-6}
        & \scriptsize{UAV} & \scriptsize{PID} & \scriptsize{PIC} & \scriptsize{PID} & \scriptsize{PIC} \\
        \hline \hline
        \multirow{3}*{\scriptsize{Scen. 1}} & \scriptsize{main} & \scriptsize{17.71 m} & \scriptsize{\textbf{16.34 m}} & \multirow{3}*{\scriptsize{\textbf{27\%}}} & \multirow{3}*{\scriptsize{30.25\%}} \\
        \cline{2-4}
        & \scriptsize{aux. 1} & \scriptsize{25.71 m} & \scriptsize{\textbf{20.57 m}} & & \\
        \cline{2-4}
        & \scriptsize{aux. 2} & \scriptsize{32.76 m} & \scriptsize{\textbf{22.06 m}} & & \\
        \hline
        \multirow{3}*{\scriptsize{Scen. 2}} & \scriptsize{main} & \scriptsize{17.51 m} & \scriptsize{\textbf{14.74 m}} & \multirow{3}*{\scriptsize{4.67\%}} & \multirow{3}*{\scriptsize{\textbf{0}}} \\
        \cline{2-4}
        & \scriptsize{aux. 1} & \scriptsize{17.76 m} & \scriptsize{\textbf{16.31 m}} & & \\
        \cline{2-4}
        & \scriptsize{aux. 2} & \scriptsize{23.55 m} & \scriptsize{\textbf{18.89 m}} & & \\
        \hline
        \multirow{3}*{\scriptsize{Scen. 3}} & \scriptsize{main} & \scriptsize{22.53 m} & \scriptsize{\textbf{19.59 m}} & \multirow{3}*{\scriptsize{5.25\%}} & \multirow{3}*{\scriptsize{\textbf{1.5\%}}} \\
        \cline{2-4}
        & \scriptsize{aux. 1} & \scriptsize{27.31 m} & \scriptsize{\textbf{22.44 m}} & & \\
        \cline{2-4}
        & \scriptsize{aux. 2} & \scriptsize{\textbf{22.58 m}} & \scriptsize{24.98 m} & & \\
        \hline
        \multirow{3}*{\scriptsize{Scen. 4}} & \scriptsize{main} & \scriptsize{23.07 m} & \scriptsize{\textbf{22.17 m}} & \multirow{3}*{\scriptsize{34.5\%}} & \multirow{3}*{\scriptsize{\textbf{12.75\%}}} \\
        \cline{2-4}
        & \scriptsize{aux. 1} & \scriptsize{48.16 m} & \scriptsize{\textbf{29.29 m}} & & \\
        \cline{2-4}
        & \scriptsize{aux. 2} & \scriptsize{\textbf{23.07 m}} & \scriptsize{27.32 m} & & \\
        \hline
        \multirow{3}*{\scriptsize{Scen. 5}} & \scriptsize{main} & \scriptsize{55.20 m} & \scriptsize{\textbf{54.67 m}} & \multirow{3}*{\scriptsize{21.44\%}} & \multirow{3}*{\scriptsize{\textbf{6.33\%}}} \\
        \cline{2-4}
        & \scriptsize{aux. 1} & \scriptsize{\textbf{55.74 m}} & \scriptsize{62.68 m} & & \\
        \cline{2-4}
        & \scriptsize{aux. 2} & \scriptsize{\textbf{55.20 m}} & \scriptsize{57.04 m} & & \\
        \hline
    \end{tabular}
\end{table}

\section{Discussions} \label{sec:discussion}

In this section, we discuss and compare the outcome of the control approach based on the results given in Section~\ref{sec:comparison_results}. 
From Table~\ref{tab:fov_intersection_time_LoS_comparison}, our proposed control approach allows having a view of the main UAV without occlusions for a larger amount of time.
The gain that can arrive to $44.00 \%$, especially high in scenarios 2 and 3 where the obstacles are more dangerous because closer to the trajectories of the UAVs.
Likewise, based on Figure~\ref{fig:fov_intersection_volume}, we observe that only in scenario 1 there is a significant amount of volume intersecting with LoS obstacle avoidance task (blue curve), because of the wider configuration of the obstacles. Nevertheless, the amount is negligible in the other four scenarios.
With regard to the comparison with and without distance task, from Figure~\ref{fig:robots_distance}, one can notice that the additional task allows in general a better performance.
In fact, from Table~\ref{tab:distance_errors} the maximum errors are usually lower.
Only in three cases the maximum errors are lower without the distance task: this happens because the contribution of the acceleration (control input) in THC to maintain the distance, generates an oscillatory behavior, and therefore the peaks can be higher in a few cases (see Figure~\ref{fig:robots_distance}). 
However, the average errors are always lower than or equal to the cases when the distance task is removed.
Finally,
from Table~\ref{tab:PIC_comparison}, we can see that in general PIC produces more optimal trajectories than PID, i.e., a shorter path to the goal, of up to $39.58 \%$. Only in 4 cases over 15, the trajectories are not shorter, and this happens because there are no obstacles directly involved in the trajectory of the second auxiliary UAV, therefore with PID it can be shorter since using PIC or PID produces a similar result.
From Table~\ref{tab:PIC_comparison}, we can also see that the use of PIC also provides
smaller intersections of the FoV with obstacles. 
Only in the first scenario the intersection is smaller without PIC, and the reason is that the amount of time with intersections of the two auxiliary UAVs overlaps more.

\section{Conclusions and topics for future research} 
\label{sec:conclusion}
In this paper, we have presented a THC approach for a multi-UAV system that integrates the best viewpoints for monitoring a main UAV performing a task by other auxiliary UAVs, a LoS obstacle avoidance approach for keeping an obstacle-free view, and PIC for optimal trajectory planning. 
Motivated by the XPRIZE Wildfire challenge \cite{xprize_wildfire} for fire suppression by a multi-UAV system, our approach can be used in any other application.
In the future, GPU acceleration for PIC should be used to run our approach in real-time.
We also plan to perform experiments in real-life with UAVs.
When dealing with obstacles that have a non-spherical shape, spheres can be used to approximate the 3D volume of the obstacles, so that they can be considered as multiple spheres. The initially known position, size and number of obstacles can be treated as unknown in the future.

\section{Acknowledgment}

This work was supported in part by the TU Delft AI Labs \& Talent programme and by the NWO Talent Program Veni project ``Autonomous drones flocking for search-and-rescue'' (18120), which has been financed by the Netherlands Organisation for Scientific Research (NWO). 
University of Texas at Austin has collaborated in this work through the project within the XPRIZE Wildfire competition.

\bibliographystyle{./IEEEtran}
\bibliography{IEEEabrv,./bibliography}

\end{document}